\documentclass[showpacs,superscriptaddress,prl,twocolumn,tightenlines,balancelastpage,10pt,a4paper]{revtex4}%
\usepackage{graphicx}
\usepackage{graphicx}%
\begin{document}
\title{Beamsplitting attack to the revised KKKP protocol and a possible
solution}

\author{Xiang-Bin Wang}
\affiliation{ Imai Quantum Computation and Information Project,
ERATO-SORST, JST, Daini Hongo White Building 201, 5-28-3, Hongo,
Bunkyo, Tokyo 113-033, Japan}

\author{Qiang Zhang}
\address{Hefei National Laboratory for Physical Sciences at
Microscale \& Department of Modern Physics, University of Science
and Technology of China, Hefei, Anhui 230026, P.R. China}
\address{ Physikalisches Institut, Universit\"{a}t Heidelberg,
Philosophenweg 12, 69120 Heidelberg, Germany}

\author{Yu-Ao Chen}
\address{ Physikalisches Institut, Universit\"{a}t Heidelberg,
Philosophenweg 12, 69120 Heidelberg, Germany}

\author{Won-Young Hwang}
\address{Department of Physics Education, Chonnam National
University, Kwangjoo 500-757, Republic of Korea}

\author{Myungshik Kim}
\affiliation{School of Mathematics and Physics, Queen's
University, Belfast BT7 1NN, United Kingdom}

\author{Jian-Wei Pan}
\address{Hefei National Laboratory for Physical Sciences at
Microscale \& Department of Modern Physics, University of Science
and Technology of China, Hefei, Anhui 230026, P.R. China}
\address{ Physikalisches Institut, Universit\"{a}t Heidelberg,
Philosophenweg 12, 69120 Heidelberg, Germany}

\begin{abstract}
\thispagestyle{empty} We show that the revised KKKP protocol
proposed by Kye and Kim [Phys. Rev. Lett. {\bf 95}, 040501(2005)]
is still insecure with coherent states by a type of beamsplitting
attack. We then further revise the KKKP protocol so that it is
secure under such type of beamsplitting attack. The revised scheme
can be used for not-so-weak coherent state quantum key
distribution.
\end{abstract}

\pacs{03.67.-a, 03.67.Dd, 03.67.Hk}

 \maketitle
Quantum key distribution (QKD) can help two legal parties (Alice and Bob) to
accomplish unconditionally secure communications which is an
impossible task by any classical method \cite{gisin}. The security
of QKD is guaranteed by known principles of quantum mechanics
\cite{wooters,ekert,shor} rather than the assumed computational
complexity in classical secure communication. It is one of the
most promising applications of quantum information science and the
gap between theory and practice has become narrower.

Single-photon QKD was the first being theoretically investigated
\cite{BB84}, experimentally realized \cite {Bennet} and proved to be
secure \cite{shor}. However, the imperfect single-photon source
leads the QKD protocol vulnerable to the photon number splitting
(PNS) attack \cite{PNS} which limits its application in practice.
Therefore, several revised protocols have been proposed using
weak-coherent states \cite{Decoy01,sarg04,Decoy02,Decoy03}.  In particular, the classical process of basis reconciliation may be an important source of information leakage to an eavesdropper, known as Eve~\cite{sarg04}.  Very
recently Kye et al. \cite{kye0} proposed a blind polarization,
where the sender and the receiver share key information by
exchanging qubits with arbitrary polarization angles without basis
reconciliation. (There is another protocol where basis
reconciliation is not necessary \cite{hwa98}.) The so-called KKKP protocol, named after the inventors' initials~\cite{kye0},  was
thought to be secure even when a key is embedded in a not-so-weak
coherent-state pulse because only randomly polarized photons are exchanged as another important advantage.  The KKKP protocol has generated considerable interests~\cite{qiang,bergou}

Despite its advantages, the KKKP has been found vulnerable to impersonation attacks, due to the fact that a key has to travel three times between the legitimate users.  As shown in \cite{qiang}, the original KKKP protocol \cite{kye0}  is insecure {\em even if} because of a mathematical loophole even for single-photon keys. The loophole has then been immediately filled up~\cite{kye0} and the protocol has been made secure for single-photon keys.  However, the strength of the KKKP protocol lies in the possibility to use coherent-state keys of reasonable intensity.  If Alice is limited to a single
photon source, the protocol seems to be rather
inefficient compared with the prior art standard protocols. In this Letter, we show the revised KKKP protocol is still insecure for coherent-state keys, and furthermore, we give a solution, which is robust against impersonation attacks.

Consider the revised KKKP protocol \cite{kye1}. (Here we slightly simplify the protocol to perform $(\pi/2)$ rotations, rather than $(\pi/4)$ rotations.) : {\bf K1.} Alice sends Bob two coherent pulses with
polarization angles $\theta_0,\theta_1$. (For the ease of
presentation, we shall use subscripts ``0, 1'' rather than
``1,2''.) {\bf K2.} After reception, Bob applies random shuffling
$U_y[\phi + {s_0} (\pi/2)]\otimes U_y[\phi + {s_1}(\pi/2)]$ where
$s_i$ is randomly chosen from $ \{0,1\}$ and $(i =0,1)$. Bob
sends the two pulses back to Alice. {\bf K3.} Upon reception,
Alice applies $U_y[-\theta_0+ {k}(\pi/2)]\otimes U_y[-\theta_1+
({k}\oplus1)(\pi/2)]$ and $k\in \{0,1\}$ is the key bit. Alice
blocks one pulse and sends the other one back to Bob. The
polarization angle of the surviving pulse is $\phi+ (s_b \oplus
k\oplus b) (\pi/2)$, and $b$ is the blocking factor for Alice to
send out pulse 0 (b=0) or pulse 1 (b=1). {\bf K4.} Bob applies
$U_y(-\phi)$ to the only pulse he receives and measures the
polarization angle. The measurement outcome reveals the value
$l=s_b\oplus k\oplus b $. Alice announces $b$ and Bob uses
$k=l\oplus b\oplus s_b$ as the secret bit shared with Alice.

We show our attack by two arguments.  {\bf  A:} If Eve knows
the parity value, $s_0\oplus s_1$, she can attack the protocol
successfully.  { \bf B:} There is indeed a way for Eve to know the
value $s_0\oplus s_1$ without causing any disturbance to Alice or
Bob's detection. The {\bf B}
can be expected from the following fact:  When $s_0 \oplus
 s_1 =0$, the protocol reduces to the single pulse KKKP protocol
\cite{kye0} actually, because Bob shuffles both qubits by the same
degree. When $s_0 \oplus s_1 =1$, the protocol also reduces to the
(double pulse) KKKP protocol \cite{kye0}, because Bob shuffles the
qubits always differently. In both cases, the protocol can be
successfully attacked by Eve. The former and the latter cases are
dealt with in Refs. \cite{kye0} and \cite{qiang}, respectively.

Now we show {\bf A} with an impersonation attack where to
Alice, Eve pretends herself to be Bob, while to Bob, Eve pretends
herself to be Alice in their quantum communication channel. We assume
that Eve does not  attack the classical communication between
Alice and Bob.
\\
{\bf Protocol A:}  {\bf A1.} After K1, Eve intercepts both pulses
from Alice and stores them in set $E1$. Meanwhile, Eve prepares
two coherent pulses by herself with polarization angles
$\theta_0',\theta_1'$. {\bf A2.} After K2, Eve intercepts both
pulses from Bob. After the treatment in the subprotocol {\bf As},
Eve stores the remaining pulses intercepted from Bob in set $E2$.
Suppose with subprotocol {\bf As}, Eve now knows the value
$s_0\oplus s_1$.
{\bf A3.} If $s_0\oplus s_1 =0$, Eve rotates the polarization of
pulses in set $E1$ by $U_y [\phi'+ s_0'(\pi/2)]\otimes U_y [\phi'+
s_0'(\pi/2)] $ and sends them to Alice. Note that $s_0'$ is set by
Eve herself. After K3, Eve intercepts the only pulse from Alice
and measures its polarization after a rotation of $U_y(-\phi')$.
The outcome reveals $l'=k\oplus s_0'$.
 Since $s_0'$ is set by Eve herself, Eve knows the value
$k$ already. Eve rotates pulse 0 in set $E_2$ by $U_y(-\theta_0' +
k (\pi/2))$ and sends it to Bob. As Alice has expected in the
protocol, Bob will obtain $l=k\oplus s_b$ for sure after he
measures the polarization.
If $s_0\oplus s_1 =1$, Eve rotates the polarization of pulses in
set $E1$ by $U_y [\phi'+ {s_0'}(\pi/2)]\otimes U_y [\phi'+
(s_0'\oplus 1)(\pi/2)] $ and sends them to Alice. After K3, Eve
intercepts the only pulse from Alice and measures its polarization
after a rotation of $U_y(-\phi')$. As one may easily see, the
outcome is simply $l'=s_0'\oplus k \oplus b$. Since $s_0'$ is set
by herself, Eve already knows the value $k \oplus b$. Eve
rotates pulse 0 in set $E_2$ by $U_y(-\theta_0' + (k \oplus
b)(\pi/2))$ and sends it to Bob. Bob will for sure obtain $l=k
\oplus b \oplus s_b$, as Alice has expected in the protocol. Since
$b$ is announced later, Eve can obtain $k$ by $k=(k\oplus b)
\oplus b$.

Protocol {\bf A} shows that Eve may have full information about
Bob's result without causing any noise, if she knows the value
$s_0\oplus s_1$. We now show that she can indeed know this by {\bf
subprotocol As}: Consider Fig. 1. After K2, Bob sends two pulses
$B0$ and $B1$ back to Alice. Eve intercepts them and splits each
of them by a beamsplitter. To know the value of $s_0\oplus s_1$ is
simply to know whether the two pulses(defined as $E0$ and $E1$)
have the same polarization angle. Eve first observes whether each
pulse contains at least one photon by a quantum non-demolition
measurement. If yes, she takes one photon from each pulse and then
guides them to a 50:50 beamsplitter after rotating the
polarization of pulse $E0$ and pulse $E1$ by
$-\theta_0,-\theta_1$, respectively. As it has been well known
\cite{pan}, if the polarization of two input photons are the same,
one output beam must be vacuum. Therefore if she observes one
photon on each output ports, she concludes that $s_0\oplus s_1=1$,
otherwise, the result is inconclusive. Before guiding pulses $E0$
and $E1$ into a 50:50 beamsplitter, Eve may choose to rotate the
polarization of beam $E0$ by $\pi/2$. In such a case, if she
observes one photon on each output ports, she concludes that
$s_0\oplus s_1=0$, otherwise, the result is inconclusive. Also, if
initially $E0$ or $E1$ is vacuum, the result is inconclusive. If
the result is inconclusive, Eve intercepts and discards everything
from Alice and Bob in protocol {\bf A}. If the result is
conclusive, Eve continues her protocol {\bf A} with the exact
information of $s_0\oplus s_1$. (After subprotocol {\bf B}, the
pulses $E0$ and $E1$ are consumed already and Eve shall store
pulses 0 and 1 in set $E2$ for protocol {\bf A}.)  This attack
protocol may sound plausible.  However, there is a limitation for
it due to a high probability of inconclusive events. The total of
75\%, Eve gets inconclusive events, and only 25\% of the case can
be used.  The best strategy for Eve is to prepare each pulses of
$B0$ and $B1$ in photon-number eigenstates containing at least 2
photons.  In this case, only with the probability of 25\%, Bob
will receive any photons.  Alice and Bob know in advance the bit
rate, depending on their channel efficiency and coherent pulse
amplitude Alice initially prepares. In \cite{kye0}, when the
channel efficiency is as low as $\eta^2=0.5$ and the initial
amplitude of Alice's pulse is $2.83$, the successful detection of
a photon by Bob is calculated as 63.5\%.  Thus 25\% is too low to
be unnoticed by Alice and Eve. If Eve prepares $B_0$ and $B_1$ in
coherent states of amplitude $\gamma$, the problem will become
more serious as the coherent states already have non-zero
probability of there being no photons.

Here, we confirm that what we discussed above is the optimum
discrimination of the two qubits in parallel or anti-parallell
polarizations (See \cite{barnett}). Eve can have more setups to
measure $s_0\oplus s_1$. For example, Eve increases the photon
numbers of $B0$ and $B1$ and take two set of single photons from
the pulses returning from Bob.  She then have two beam splitter
setup: one setup to give a conclusive event from the two qubits
being in the same polarization and the other from them being in
the anti-parallel polarization.  This can however take enormous
resources from Eve.  Of course, Eve can use spy pulses in a
singlet state to find if Bob's operations are same or orthogonal
for the two pulses but we stick to only impersonation attack as
this will suffice our needs as seen later.

\begin{figure}\includegraphics[width=2in]{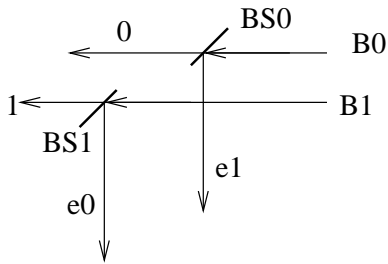}
 \caption{ Eve can know the value $s_0\oplus
s_1$ by detecting beam $E0$ and $E1$ appropriately. BS0 and BS1
are two identical beam splitters. Pulses 0 and 1 are stored in set
$E2$. Beams $B0$ and $B1$ are sent by Bob after step K2.
 }
\end{figure}

Moreover, in the present form of the revised KKKP protocol, Bob
does not randomly change the intensity or phase of the two pulses
before he sends them back to Alice therefore the two pulses sent
out from Bob have the same intensity and they are phase-locked. In
such a case, we can improve it using coherent state attack and
photon number discriminator detector is not to be necessary.

If two coherent fields of the same amplitude are inputs to a beam
splitter \cite{kim}, we know that the coherent state will be
driven into only one output port. Let us assume that Eve prepares
with a relatively large intensity fields (amplitude is $\gamma$).
Eve sends them to Bob as she impersonates Alice to Bob. Upon
reception of the pulses, Bob adds rotations and shuffling factors.
As is shown in Fig. 2, receiving the pulses from Bob, Eve will
first split the pulses by a beamsplitter 1 (BS1). The coherent
fields reflected by BS1
 (amplitude $r\gamma$, where r is the reflectivity of BS1) will be used to
 measure $s0 \oplus s1$ and the transmitted field (amplitude $t\gamma$, where t is the transmissivity)
 will be stored and eventually sent back to Bob. The reflected pulses will be further
 split into two by a 50:50 beam splitter (BS2). The reflected pulses of BS2 will be used to
 measure if the two pulses are parallel to each other, while the transmitted pulses will be used to
 check if they are orthogonal. Here we only consider the orthogonal measure. The first and second are split
 by the dynamic reflector (DR) to give a time delay to the first pulse to meet the second at a 50:50 beam splitter (BS3).
  If they are of the same polarization, both pulses will be detected in photon detector 1 (PD1).
  The amplitude of the coherent field to PD1 is $\sqrt{2}r\gamma$ and if they are orthogonal,
  they do not interfere and the amplitude of coherent fields to PD1 and PD2 are both $r\gamma$.
  If Eve detects any photon, then it is due to the fact that the two were orthogonal.

Here, we remind that for a non-zero amplitude coherent field,
there is non-zero probability of no photon detected so there is also a chance of inconclusive events.  The success probability is calculated using the Poissonian nature of coherent state.
$P_{success}=(1-\mbox{e}^{-r^2\gamma^2})^2$.   Similarly, Eve
measures if two pulses are parallel to each other by rotating the
polarization angle of one of the pulses by $\pi/2$. The success
probability is again $P_{success}$. Therefore, the total probability of success is also $P_{success}$.

Eve then let the pulses go to Bob in the conclusive cases for which Bob
still has a probability of receiving zero photons as a coherent
state of $t\gamma$ will travel to Bob. The total probability of
Bob receiving any photons then is $P_{B} = P_{success}(1-\mbox{e}^{-t^2\gamma^2})$. Eve knows the bit rate so she can make $P_B$ to make it equal to the bit rate by changing  $\gamma$.  Eve then has implemented the subprotocol {\bf As} successfully without having her action noticed by Alice and Bob.
The revised KKKP protocol, with the idea  being interesting, in
its present form does not offer the security as it has been
supposed therefore a more careful investigation is needed for real
applications. Now we offer a possible solution.

\begin{figure}\includegraphics[width=2.9in]{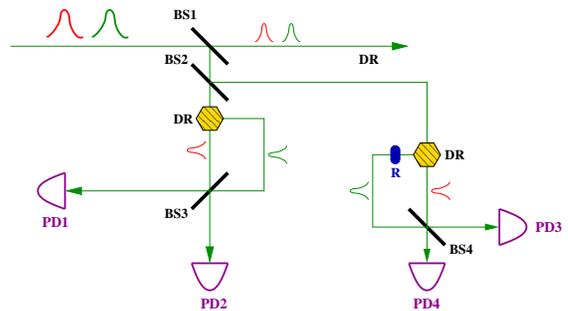}
 \caption{ Scheme for Eve to find if the two coherent pulses are
of orthogonal or parallel polarization. DR: dynamic reflector to reflect only the first pulse while the second pulse is let to transmit.  BS: beam splitters. PD: photo detectors}
\end{figure}

Consider the following protocol {\bf W}:\\
{\bf W1:} same with step K1 of KKKP protocol. {\bf W2: } Upon
reception, Bob applies random shuffling $U_y[\phi
+{s_0}(\pi/2)]\otimes U_y[\phi+\delta(\pi/4) + s_1(\pi/2)]$ where
$s_i\in \{0,1\} (i\in 0,1)$, $\delta\in\{0,1\}$ are  random
numbers. Bob sends the two pulses back to Alice. {\bf W3:}  Upon
reception, Alice decides randomly to either use the received
pulses for test or continue the protocol for  sharing a secret bit
with Bob. If she decides to continue, she applies $U_y[-\theta_0+
k(\pi/2)]\otimes U_y[-\theta_1+ k(\pi/2)]$ and $k\in \{0,1\}$ is
the key bit. Alice blocks one pulse and sends the other one back
to Bob. (Alice also detects the blocked pulse to make sure that
this is not in vacuum.) The polarization angle of the surviving
pulse is $\phi+ (s_b \oplus k \oplus b)(\pi/2)+  b \delta
(\pi/4)$, and $b$ is the blocking factor for Alice to send out
pulse 0 (b=0) or pulse 1 (b=1). Alternatively, Alice may decide to
consume the pulses from Bob to inspect the presence of Eve who
tries to detect the blocking factor $b$ by sending her two
different pulses of photon-number eigenstates (Eve cannot use coherent pulses of different amplitude because it is not possible to tag them as two coherent states are not orthogonal). In order to do the test, she can use the coherent-state comparison technique as used by Eve described above.  In this case, when the two coherent-state pulses of the same amplitudes should be driven to only one output port while the two photon-number eigenstates of different photon numbers would appear in the both output ports.  Of course, there also is some non-zero probability of two photon-number eigenstates to appear in only one output port.  Alice can check the probability of the all-or-nothing events and if it is less than desired she will have to abolish the keys assuming there having been Eve in the middle.
{\bf W4.} Bob applies
$U_y[-\phi-\Delta(\pi/4)]$ to the only pulse he receives and
measures the polarization angle.  Pulse 1 is randomly chosen from 0 or
$\delta$. If $\Delta$ happens to be equal to $b\delta$, the
measurement outcome reveals the value $l=k\oplus b\oplus s_b$.
Alice announces $b$ and Bob uses $k=l\oplus b\oplus s_b$ as the
secret bit shared with Alice if $\Delta=b\delta$. If
$\Delta\not=b\delta$, they discard the data. {\bf W5.} They run
the above program for $N$ times. {\bf W6.} Alice announces which
times she has used the pulses for test in step W3. Bob announces
the values of $\delta$ and $s_0,s_1$ he has chosen for those
times. Alice only needs to consider those testing results where
her rotation angle $\omega(\pi/4)$ happens to be equal to $[
2(s_0+ s_1 )- \delta ](\pi/4)$. If they are all single-clickings,
she judges that there is no Eve, otherwise, she aborts the
protocol. {\bf W7.} Alice and Bob also compare some of their bits
through classical communications to see whether they have indeed
shared the same key.

In the protocol above, there are two different types of error
tests. Step $\bf W6$ is to test  whether there is Eve who tries
to detect the blocking factor $b$ by sending Alice two different
pulses. $\bf W7$ is to test whether there is Eve who tries to
know $k$ or equivalently, $s_0\oplus s_1$. Given protocol W, Eve
cannot use the subprotocol B to obtain $s_0\oplus s_1$ without
causing any noise. Say, no matter how she rotates beam E0 or E1,
any value of $s_0\oplus s_1$ can cause the event of two-fold
clicking.  Without the exact information about $s_0\oplus s_1$ at
that time, she cannot continue protocol {\bf A} because she would cause
noise in Bob's key and this may be detected in step {\bf W7}.

In summary, we have shown that the revised KKKP
protocol\cite{kye1} in its present form fails under a type of
beamsplitter attack. We also propose a protocol which is robust against any impersonation attack.

\begin{acknowledgments}
MSK thanks Erika Andersson for discussions on the optimal distinguishability ot two qubits.
This work was supported by the NNSF of China, the CAS, the PCSIRT
and the National Fundamental Research Program, the Marie Curie
Excellent Grant of the EU, the Alexander von Humboldt Foundation
and the Deutsche Telekom Stiftung. This work was also supported by
the Korea Research Foundation Grant funded by the Korean
Government (MOEHRD) (KRF-2005-003-C00047).
\end{acknowledgments}


\begin{thebibliography}{99}
\bibitem {gisin} N. Gisin, G. Ribordy, W. Tittel, and H. Zbinden, Rev. Mod. Phys. \textbf{74} 145(2002).

\bibitem {wooters} W. K. Wooters and W. H. Zurek, Nature(London), \textbf{299}, 802(1982).

\bibitem {ekert} A. K. Ekert, Phys. Rev. Lett., \textbf{67}, 661(1991).

\bibitem {shor} P. W. Shor and J. Preskill, Phys. Rev. Lett., \textbf{85}, 441(2000).

\bibitem {BB84} C. H. Benneett and G. Brassard, in Proceedings of the IEEE International Conference on Computes, Systems
and Signal Processing, p. 175(1984).

\bibitem {Bennet} C. H. Bennett, F. Bessette, G. Brassard, L. Salvail, and J. Smolin,  J. of Cryptology \textbf{5},
3(1992).

\bibitem {PNS}G. Brassard, N. Lutkenhaus, T. Mor, and B. C. Sanders, Phys. Rev. Lett., \textbf{85}, 1330(2000)

\bibitem {Decoy01}W.-Y. Hwang, Phys. Rev. Lett., \textbf{91}, 057901(2003).

\bibitem{sarg04} V. Scarani, A. Acin, G. Ribordy, N. Gisin, Phys. Rev. Lett., \textbf{92},
057901(2004).


\bibitem {Decoy02}X. B. Wang, Phys. Rev. Lett., \textbf{94}, 230503 (2005).

\bibitem {Decoy03}H. K. Lo, X.-F. Ma and K. Chen, Phys. Rev. Lett., \textbf{94}, 230504
(2005).

\bibitem{kye0} W.-H. Kye, C. Kim,  M. S. Kim, and Y. J. Park, Phys. Rev.
Lett. \textbf{95}, 040501(2005).

\bibitem{hwa98} W.-Y. Hwang, I.G. Koh, and Y.D. Han, Phys. Letts.
\textbf{244}, 489 (1998).

\bibitem{qiang} Q. Zhang, X.-B. Wang, Y.-A. Chen, W.-Y. Hwang, T. Yang, and
J.-W. Pan, Phys. Rev. Lett. \textbf{96}, 0789071(2006).


\bibitem{bergou} J. A. Bergou and L. B. Kish, quant-ph/0509097; S. Kak, Found. Phys. Lett.  {\bf 19}, 293 (2006).


\bibitem{kye1} W.-H. Kye and  M. S. Kim.,  Phys. Rev. Lett.
\textbf{96}, 040502(2006).

\bibitem{pan} For example, D. Bowmeester, J. W. Pan, K. Mattle, M. Eibl, H.
Weinfurter and A. Zeilinger, Nature(London), \textbf{390},
575(1997).

\bibitem{barnett} S. M. Barnett, A. Chefles and I. Jex, Phys. Lett A {\bf 307}, 189 (2003); E. Andersson, I. Jex and S. M. Barnett, J. Phys. A: Math. Gen. {\bf 36}, 2325 (2002)

\bibitem{kim} M. S. Kim, W. Son, V. Bu\v{z}ek and P. L. Knight, Phys.
Rev. A \textbf{65}, 032323 (2002).


\end{thebibliography}
\end{document}